\documentclass[aps, pra,twocolumn, showpacs,superscriptaddress,10pt,floatfix]{revtex4}
\usepackage[english]{babel}
\usepackage[latin1]{inputenc}
\usepackage{hyperref}
\usepackage{amsmath, amssymb, amsfonts}
\usepackage{amssymb,xcolor}
\usepackage{graphicx}
\usepackage{exscale}
\usepackage{latexsym}
\usepackage{cases}
\usepackage{epstopdf}
\usepackage{bbm}
\usepackage{subfigure}
\usepackage{dsfont}
\usepackage{color}

\newcommand{\ket}[1]{|#1\rangle}
\newcommand{\bra}[1]{\langle #1|}
\newcommand{\Tr}{\mathrm{Tr}}

\newcommand{\n}{\nonumber\\}

\newcommand{\p}{\partial}

\newcommand{\ex}[1]{\langle #1\rangle}

\allowdisplaybreaks

\begin{document}
\title{Quantum-Enhanced Metrology in  Cavity Magnomechanics}

%
%
%
%


\author{Qing-Kun Wan}
\affiliation{Innovation Academy for Precision Measurement Science and Technology, Chinese Academy of Sciences, Wuhan 430071, China}
\affiliation{University of Chinese Academy of Sciences, Beijing 100049, China}
%

\author{Hai-Long Shi}
\email{hl\_shi@yeah.net}
\affiliation{ Innovation Academy for Precision Measurement Science and Technology, Chinese Academy of Sciences, Wuhan 430071, China}
\affiliation{University of Chinese Academy of Sciences, Beijing 100049, China}

\author{Xi-Wen Guan}
\email{xiwen.guan@anu.edu.au}
\affiliation{Innovation Academy for Precision Measurement Science and Technology, Chinese Academy of Sciences, Wuhan 430071, China}
\affiliation{NSFC-SPTP Peng Huanwu Center for Fundamental Theory, Xi'an 710127, China}
\affiliation{Hefei National Laboratory, Hefei 230088 Chia}
\affiliation{Department of Fundamental and Theoretical Physics, Research School of Physics,
Australian National University, Canberra ACT 0200, Australia}


\date{\today}

\begin{abstract}

Magnons, as fundamental quasiparticles emerged in elementary spin excitations, hold a big promise for  innovating quantum technologies  in information coding and processing.
By establishing the exact relation between Fisher information and entanglement in partially accessible metrological schemes, 
we rigorously prove  that bipartite entanglement plays a crucial role during the dynamical encoding process.
However, the presence of an entanglement during the measurement process unavoidably reduces the ultimate measurement precision.
These findings are verified in an experimentally feasible cavity magnonic system engineered for detecting a weak magnetic field by performing precision measurements  through  the cavity field.
Moreover, we further demonstrate that within a weak coupling region, measurement precision can reach  the Heisenberg limit. 
Additionally, quantum criticality also enables us to enhance measurement precision in a strong coupling region.

\end{abstract}
\pacs{03.65.Ta, 06.20.Dk, 03.67.-a, 75.10.Pq}

\maketitle
\emph{Introduction.}--In contrast to  classical estimation
theory, quantum metrology seeks a higher-precision estimation of some fundamental physical quantities by using various quantum resources, such as  squeezed light\,\cite{Caves80,Caves81}, entanglement \cite{Giovannetti06,Pezze18,Hyllus12,Toth12,Li13,Ren21,Su20}, steering\,\cite{Yadin21,Frowis19,Gianani22}, nonlocality\,\cite{Niezgoda21}, and discord\,\cite{Girolami14,Girolami15,Sone19},  etc. 
In this scenario, quantum information science combined with condensed matter physics strikingly deepens our  understanding of quantum features of quasiparticles and opens a new avenue of  implementing quantum-enhanced metrology by virtue of quasiparticles\,\cite{Jurcevic14,Alvaredo18,Laflorencie16,Venema16,Wen17,Wen19,Chen10,Eisert15,Calabrese07,Scarani04},
for example, quantum metrology at  quantum criticality\,\cite{Venuti07,Zanardi08,Gu08,Albuquerque10,Frerot18,Garbe20,Montenegro21,Chu21,Liu21,Gu10,Ilias22,Yang23,Candia23} and by  dynamic structure factor\,\cite{Krammer09,Hauke16}. 

Magnons, as the collective magnetic excitations of the magnetically ordered states in interacting spins, have received increasing attention due to their capability for carrying, transporting and processing  quantum  information\,\cite{Chumak15,Lenk2011,Yuan22}.
Appealing features of magnons include their stability at room temperature, high spin density, no Ohmic losses, and fine tunability of spin orientations. 
Promising potential applications have stimulated recent research on cavity magnonics, especially on magnon-photon entanglement  and magnon-magnon entanglement\,\cite{Rameshti22,Li20,Harder21,Li18,Zhang16,Yu20,Yuan20,Huebl13,Quirion20,Wolski20,Soykal10,Elyasi20,Zhang19,Azimi21,Yuan20-2}, 
and  the role of entanglement in quantum batteries\,\cite{QB1,QB2,QB3}.
For quantum metrology, entanglement is known to provide enhanced sensing precision in  global parameter estimations requiring also global measurements\,\cite{Giovannetti06,Hyllus12,Toth12}.
However, global accessibility is often limited in cavity magnonic systems\,\cite{Montenegro22,Montenegro20}.
Natural questions arise: 
what is the role of bipartite entanglement in partially accessible metrological schemes?
And whether quantum-enhanced metrology can be realized in such  cavity magnonic systems?  

In this Letter, we address the aforementioned questions by analyzing the influence of bipartite entanglement on the scaling of quantum and classical Fisher information (QFI and CFI), which quantify the measurement precision.
We identify two important conditions for enhancing measurement precision: encoding the estimated parameter within the covariance matrix of the partially accessible quantum state and  minimizing thermalization during the final measurement process as well.
The former is ensured by the  bipartite entanglement generated through quantum dynamics, whereas  the latter is to avoid bipartite entanglement in the measurement process.
Based on this  finding, we verify  optimal measurement precisions for different settings in the experimentally feasible cavity magnonic system \cite{Tabuchi14,soykal:PRB2010,Zhang14,Bourhill16}. 
We show that the Heisenberg limit (HL) can be achieved in  the weak coupling case with an initial squeezed magnon, whereas criticality-enhanced metrology can be realized in the strong coupling case without a need for quantum squeezing.
%
%

\begin{figure}
	\includegraphics[width=0.95\linewidth]{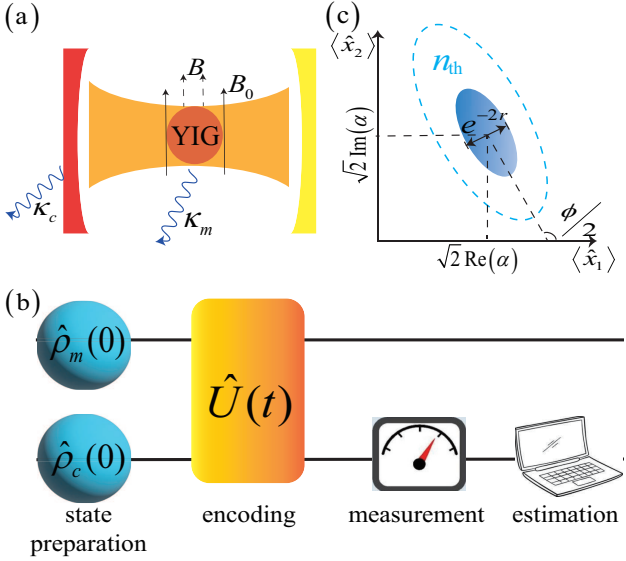}
	\caption{
(a) 
Schematic illustration of the cavity magnonic system:
A YIG sphere is placed in a cavity to sense the weak magnetic field $B$, with a bias magnetic field $B_0$ applied for magnon excitation.
$\kappa_c$ and $\kappa_m$ represent the damping rates  of the cavity and the magnon, respectively.
(b) 
Metrological scheme for  estimating the parameter $B$:
Quantum dynamics encode the estimated parameter
within the covariant matrix, while the measurement is performed on the cavity.
(c)  Illustration  of the  Wigner function of a single-mode Gaussian state in terms of displacement, squeezing, phase, and thermalization parameters $(\alpha, r, \phi, n_{\rm th})$.
}
	\label{fig:fig1}
\end{figure}

\emph{Role of entanglement in partially accessible metrological schemes}--
Partially accessible metrological schemes aim to estimate a parameter $B$ from the subsystem $\hat \rho_c\equiv\Tr_m(\hat\rho_{cm})$ where $c$ and $m$ represent two distinct subsystems, such as cavity and magnon.
Focusing on a pure state for the total system $\hat \rho_{cm}$ and assuming a Gaussian state for $\hat \rho_c$, we express $\hat\rho_c$ as $\hat \rho_c\!=\!\hat D(\alpha)\hat{S}(\zeta)\hat{\rho}_{\rm th}\hat{S}^\dagger(\zeta)\hat{D}^\dagger(\alpha)$\,\cite{FN-Gaussian} in terms of the parameters $(\alpha, r, \phi,n_{\rm th})$\,\cite{Adam95}:
\begin{flalign}\label{d-CM-2-StandForm}
&\alpha=\frac{1}{\sqrt{2}}(d_1^c+id_2^c),
\quad
r=\frac{1}{2}{\rm arcosh}\left(\frac{\Tr(\gamma^c)}{2\sqrt{\det(\gamma^c)}}\right),\n
&
n_{\rm th}=\frac{\sqrt{\det(\gamma^c)}-1}{2},\quad
\tan\phi=\left(\frac{2\gamma_{12}^c}{\gamma_{11}^c-\gamma_{22}^c}\right),
\end{flalign}
where $d^c_j\!=\!\ex{\hat x_j}$ is the displacement vector, $\gamma^c_{jk}\!=\!\ex{\{\hat x_j\!-\!\ex{\hat x_j},\hat x_k\!-\!\ex{\hat x_k} \}}/2$ is the covariance matrix, $\hat x_1\!=\!(\hat c^\dag\!+\!\hat c)/\sqrt{2}$,  $\hat x_2\!=\!i(\hat c^\dag\!-\!\hat c)/\sqrt{2}$, and $\ex{\hat A}\!=\!\Tr(\hat\rho_c \hat A)$.
The parameters $(\alpha, r, \phi,n_{\rm th})$ characterize the displacement, squeezing, phase, and thermalization for subsystem $\hat \rho_c$, respectively, see Fig.\,\ref{fig:fig1}(c).
The estimation precision provided by $\hat\rho_c$ is  the Cram\'er-Rao bound\,\cite{Rao45,Cramer46,Helstrom67,Liu20}, i.e. a lower bound 
$(\delta B)^2\!\geq\!1/F_Q(\hat \rho_c)$.
In terms of parameters $(\alpha, r, \phi,n_{\rm th})$\,\cite{FN}, we analytically derive the QFI, see  Supplemental  Material (SM)\,\cite{SM} (see also references\,\cite{Paris09,Hall15} therein).
\begin{flalign}
&F_Q(\hat\rho_c)\label{QFI-final}
\!=\!
\frac{4}{2n_{\rm th}+1}\left[\alpha'\bar{\alpha}'\cosh(2r)\!+\!\text{Re}(\bar{\alpha}'^2e^{i\phi}) \sinh(2r)\right]\n
&+\frac{n_{\rm th}'^2}{n_{\rm th}(1\!+\!n_{\rm th})}\!+\!\frac{(1+2n_{\rm th})^2}{2(1\!+\!2n_{\rm th}\!+\!2n^2_{\rm th})}\! \left[\sinh^2(2r)\phi'^2\!+\!4r'^2\right]\!,
\end{flalign}
where $\bar\alpha$ is the complex conjugate of $\alpha$ and the prime denotes derivative with respect to $B$.

If $\hat \rho_c$ does not arise from critical dynamics, it is justifiable to assume finite values for $(\alpha, r, \phi, n_{\rm th})$ and their respective derivatives.
Enhancing measurement precision primarily relies on particle number, i.e., $N_c\!\equiv\!|\alpha|^2\!+\!n_{\rm th}\!+\!(2n_{\rm th}\!+\!1)\sinh^2r$.
We first consider the case where particle number is mainly sourced by the displacement, e.g., a coherent state, $N_c\!\sim\! |\alpha|^2$.
Then, Eq.\,(\ref{QFI-final})  implies $F_Q\!\sim\! \alpha'\bar\alpha'$ leading to the shot noise limit (SNL), i.e.,  $F_Q\!\sim\! N_c$.
Here it is reasonable to  assume that the parameters and their derivatives are of the same order. 
In the second case, if the thermalized photons dominate, i.e., $N_c\!\sim \!n_{\rm th}$ then  QFI (\ref{QFI-final}) does not increase with $N_c$ since $F_Q\!\sim \!\mathcal O(1)$.  

Excluding the displacement and thermalized photons as metrological resources, quantum squeezing emerges as a vital  resource for surpassing the shot noise limit, also see the study  in \,\cite{Dodonov02,Caves81,Hollenhorst79,Walls83,McKenzie02,McKenzie04,Vahlbruch05,Vahlbruch06}.  
In the third case, assuming that quantum squeezing is dominant then $N_c\!\sim\! \sinh^2r\!\sim\! e^{2r}$ and  $F_Q\!\sim\! \sinh^2(2r)\!\sim\! e^{4r}$. 
Explicitly, it achieves HL, $F_Q\!\sim\! N_c^2$. 
To incorporate the influence of the inevitable thermalization in QFI, we assume the number of thermalized photons scales as $n_{\rm th}\!\sim\! e^{\nu r}$\,\cite{FN-0}.
Equation\,(\ref{QFI-final})  reveals that  the thermalization does not really  affect the scaling of QFI  if $n_{\rm th}$ dominates.
However, $N_c \!\sim\! e^{(2+\nu)r}$ can essentially depend on the thermalization. 
So QFI exhibits a scaling
\begin{flalign}\label{scaling-1}
F_Q\!\sim\! N_c^{4/(2+\nu)},
\end{flalign}
 which can exceed the SNL  if $\nu\!<\!2$.

Bipartite entanglement can be quantified by 
the entanglement entropy $S(t)\!=\!-\Tr[\hat\rho_c\log_2(\hat\rho_c)]$\,\cite{Vedral97}.
For Gaussian states, it reduces to\,\cite{Agarwal71}
\begin{flalign}\label{ent}
S(t)=\log_2(n_{\rm th}+1)+n_{\rm th}\log_2\left(\frac{n_{\rm th}+1}{n_{\rm th}}\right),
\end{flalign}
showing  that entanglement increases with the number of thermalized photons $n_{\rm th}$ grows.
Combining  Eqs.\,(\ref{scaling-1}) and\,(\ref{ent}), we deduce that the existence of bipartite entanglement decreases the final measurement precision. 
 Nevertheless, in order to achieve scaling law\,(\ref{scaling-1}),  the parameters are required to  be encoded effectively in the covariance matrix of the subsystem $\hat\rho_c$. 
The emergent entanglement guarantees such an  encoding from a product state, thus highlighting the significance of bipartite entanglement in the dynamical encoding process.  
This finding  establishes fundamental laws governing the role of bipartite entanglement in partially accessible Gaussian metrology.

\emph{Quantum Measurements.}--
Now we discuss how to experimentally realize the measurement precision given by the Cram\'er-Rao bound.
Based on the Gaussian measurements, the CFI is obtained as follows\,\cite{SM}  (see also reference\,\cite{Monras06} therein).
\begin{flalign}\label{CFI-final}
F_C(\hat\rho_c)
&=\frac{[n'_{\rm th}-(2n_{\rm th}+1)r']^2}{2(n_{\rm th}+1)^2}
+\frac{[n'_{\rm th}+(2n_{\rm th}+1)r']^2}{2(n_{\rm th}+1)^2}\n 
& +\frac{(2n_{\rm th}+1)^2[\phi'\sinh(2r)]^2}{4(n_{\rm th}+1)^2},
\end{flalign}
where we only considered quantum states with zero displacement.
Equation\,(\ref{CFI-final}) reveals that the scaling of CFI matches that of QFI, demonstrating that  Gaussian measurements are optimal.
Experimentally, after acquiring probabilities through Gaussian measurements, we can utilize the maximum likelihood estimator  to estimate the unknown parameter\,\cite{Rossi18}.	
Implementing Gaussian measurements involves two steps: applying a Gaussian unitary operation to the input system, which includes additional ancillary (vacuum) modes,  and performing homodyne measurements on all output modes\,\cite{Weedbrook12,Giedke02,Eisert03}.
Next, we will apply these general results  to a cavity magnonic system, exploring distinct roles of entanglement in dynamic encoding and measurement processes.

\emph{Model.}--
We consider a system consisting of a Yttrium iron garnet (YIG) sphere placed inside a microwave cavity and subjected to a static magnetic field, $B_0$\,\cite{Tabuchi14, Zhang14} see Fig.\,\ref{fig:fig1}(a).
The microwave field induces magnon excitations in the ferromagnetic YIG sphere. 
Additionally, we introduce a weak field $B$ for estimation.
Using the Holstein-Primakoff approximation\,\cite{Holstein40}, the corresponding Hamiltonian is given by\,\cite{Zhang14, FN-2}
\begin{flalign}\label{H-1}
\hat H=\omega_c\hat{c}^\dagger\hat{c}+\omega_{m}\hat{b}^\dagger\hat{b}+g(\hat c+ \hat{c}^\dagger) (\hat{b}+\hat{b}^\dag),
\end{flalign}
where $\hat c^\dag$ $(\hat b^\dag)$ and $\hat c$ $(\hat b)$ are creation and annihilation operators for the cavity photon (magnon) at frequency $\omega_c$ [$\omega_m\!=\!\mu(B_0\!+\!B)]$, respectively. 
The gyromagnetic ratio $\mu$ is set to 1 and $g$ is the coupling strength.
The Hamiltonian (\ref{H-1}) applies to the ferromagnetic YIG sphere only when 
$g\!<\!g_c\!\equiv\!\sqrt{\omega_c\omega_m}/2$\,\cite{Emary03}. 
Beyond the critical point $g_c$, the system remains in the super-radiant phase.

The metrological scheme is shown in Fig.\,\ref{fig:fig1}(b),  where the initial magnon-cavity state is prepared in a product state $\hat \rho_{cm}(0)\!=\!\hat \rho_c(0)\otimes\hat \rho_m(0)$. 
Over time, the information of the weak field $B$ becomes encoded in the state $\hat\rho_{cm}(t)\!=\!\exp(-iHt)\hat\rho_{cm}(0)\exp(iHt)$.
At  time $t_*$, we perform  Gaussian measurements on the cavity state $\hat\rho_c(t_*)\!=\!\Tr_m[\hat\rho_{cm}(t_*)]$ and estimate the value of $B$ from the measurement results.

\emph{Far away from critical dynamics.}--
When $|\omega_c-\omega_m|\ll 1$ and $g\ll g_c$, we can employ the rotating wave approximation (\ref{H-1}) to rewrite the Hamiltonian as 
\begin{flalign}\label{H-RWA}
\hat H=\omega_c\hat{c}^\dagger\hat{c}+\omega_{m}\hat{b}^\dagger\hat{b}+g\left( \hat{c}^\dagger\hat{b}+\hat{c}\hat{b}^\dagger\right).
\end{flalign}
Its dissipative dynamics is described by the quantum Langevin equation
\begin{flalign}\label{Leq-RWA}
&\p_t\hat{c}(t)=-i\omega_c \hat{c}(t)-ig\hat{b}(t)-\frac{\kappa_c}{2}\hat{c}(t)+\sqrt{\kappa_c}\hat{c}_{\rm in}(t),\n
&\p_t\hat{b}(t)=-i\omega_{m} \hat{b}(t)-ig\hat{c}(t)-\frac{\kappa_m}{2}\hat{b}(t)+\sqrt{\kappa_m}\hat{b}_{\rm in}(t),
\end{flalign}
where $\kappa_c$ and $\kappa_m$ denote the damping rates of the cavity mode and the magnon mode, respectively.
The input noises are described by the annihilation operators $\hat c_{\rm in}$ and $\hat b_{\rm in}$, satisfying
$\langle\hat{c}^\dagger_{\rm in}(t_1)\hat{c}_{\rm in}(t_2)\rangle\!=\!n_{c}\delta(t_1\!-\!t_2)$ and $\langle\hat{b}^\dagger_{\rm in}(t_1)\hat{b}_{\rm in}(t_2)\rangle\!=\!n_{m}\delta(t_1\!-\!t_2)$. 
Here the values of $n_c$ and $n_m$ are subject to external  environment  thermal noise. 
For simplicity,  we assume $\kappa_m\!=\!\kappa_c$ and $n_m\!=\!n_c$.

\begin{figure}[t]
	\includegraphics[width=\linewidth]{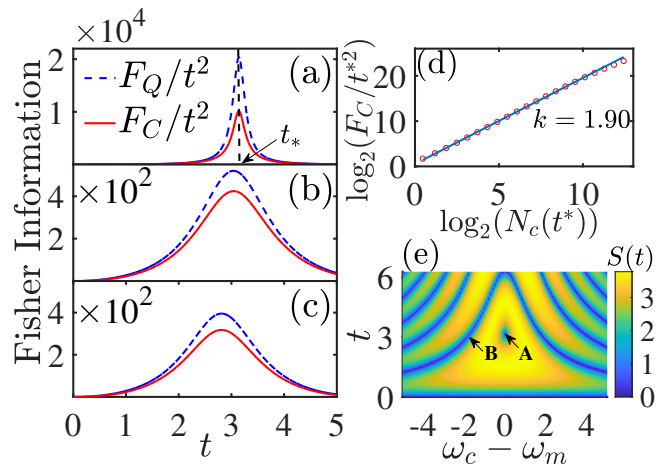}
	\caption{
Time evolution of the time-rescaled QFI $F_Q/t^2$ (blue dotted line) and CFI $F_C/t^2$ (red line)  
in (a) resonance case $\omega_c\!=\!\omega_m\!=\!2,\kappa_c\!=\!n_c\!=\!0$,
(b) dissipated case $\omega_c\!=\!\omega_m=2, \kappa_c\!=\!0.001, n_c\!=\!30$, 
and (c) off-resonance case $\omega_c\!=\!2, \omega_m\!=\!2.5$, $\kappa_c\!=\!n_c\!=\!0$.
(d) Scaling law of the CFI ($F_C\!\sim\! N_c^k$ with $ k\!=\!{1.90}$) for the parameter setting used in  the experiment \cite{Bourhill16}:   $\omega_c\!=\!\omega_m\!=\!15.506\times 2\pi$ GHZ, $g\!=\!7.11\!\times\! \pi$ GHZ, and $\kappa_c\!=\!\kappa_m\!=\!1.029\!\times\! \pi$ MHZ.
$t_*$ denotes the time of the CFI peak and $N_c(t_*)$ is the number of photons  at that moment.
(e) Time evolution of entanglement $S(t)$ versus the detuning $\omega_c\!-\!\omega_m$.
	}
	\label{fig:fig2}
\end{figure}

The initial state is chosen with the cavity in the vacuum state $\ket{0}$ and the magnon in a squeezed vacuum state $\hat\rho_m(0)\!=\!\hat{S}(r_0)\ket{0}\bra{0}\hat{S}^\dagger(r_0)$,
which  can be generated via parametric pumping\,\cite{Li19,Demokritov06,Zhang21}. 
The quantum Langevin equation (\ref{Leq-RWA}) is analytically solved and
the evolution of $\hat\rho_c(t)$ can be divided into two processes\,\cite{SM}:
\begin{flalign}
&\hat \rho_c(0)\xrightarrow[\rm P1]{\rm no\, noise} \hat\rho_{\rm in}(t) \xrightarrow[\rm P2]{\rm  noise} \hat \rho_c(t),\label{process}
\end{flalign} 
where $\hat \rho_{\rm in}(t)$ denotes the cavity state in noiseless case.
During the evolution the displacement of $\hat \rho_c(t)$ is always zero.
Focusing on the corresponding covariance matrices, we find\,\cite{SM}
\begin{flalign}\label{Pro}
&{\rm P1}:\ \gamma_{\rm in}^c(t)=\xi(t)\gamma^c_{\rm sq}+[1-\xi(t)]\mathds{1}_2,\n
&{\rm P2}:\ \gamma^c(t)=\eta(t)\gamma^c_{\rm in}(t)+[1-\eta(t)](2n_c+1)\mathds{1}_2,
\end{flalign} 
where $\gamma^c_{\rm sq}$ is the covariance matrix of the squeezed state $\hat S(-r_0e^{-i(\omega_c+\omega_m)t})\ket{0}$, $r_0$ is the squeeing parameter of the  initial magnon state,  $\gamma_{\rm in}^c(t)$ and $\gamma^c(t)$ are the covariance matrices of $\hat \rho_{\rm in}(t)$ and $\hat\rho_c(t)$, respectively.
Here $\eta(t)\!=\!\exp(-\kappa_ct)$,
$\xi(t)\!=\!4g^2\sin^2(\Delta t/2)/\Delta^2$, and $\Delta\!=\!\sqrt{4g^2+(\omega_c-\omega_{m})^2}$.

In the dissipative process P2, equation\,(\ref{Pro}) implies that $\gamma^c(t)$ tends toward a thermal state $(2n_c+1)\mathds{1}_2$ since
$\lim_{t\to\infty}\eta(t)=0$. 
This process indicates the gradual dissipation of information into the external environment, resulting in measurement precision described by the QFI (CFI) being smaller than the non-dissipative case, see Fig.\,\ref{fig:fig2}(a,b). 

Unlike process P2, process P1 describes the flow of the magnetic field information between the cavity and the magnon.
At $t\!=\!0$, the cavity is only a vacuum state without any quantum resources and information.
Then the appearance of the dynamics-induced bipartite entanglement essentially  lead to the transmission of information  from the squeezed magnon state into the cavity, see Fig.\,\ref{fig:fig2}(e). 
From Eq.\,(\ref{Pro}), we observe that the cavity state   finally becomes a  squeezed state $\hat S(-r_0e^{-i(\omega_c+\omega_m)t})\ket{0}$\,\cite{Footnote} at a time $t_*\!=\!\pi/(2g)$ satisfying $\xi(t_*)\!=\!1$.
This condition  is satisfied  only for the resonant case.
It is crucial to emphasize that at this  special time $t_*$, the whole system $\hat\rho_{cm}(t_*)$ is in a non-entangled state, yet both the information $\omega_m$ and the initial squeezing resource $r_0$ have been completely transferred to the cavity part without thermalization.
 Thus, the HL-precision, ($F_C\!\sim\! N_c^{2}$), can be achieved in the absence of noise by Eq.\,(\ref{scaling-1}).
Using  experimental parameters\,\cite{Bourhill16}, we show that  the precision still remains near HL, specifically $F_C\!\sim\! N_c^{1.90}$ as shown in Fig.\,\ref{fig:fig2}(d). 
Here, the Gilbert damping of magnons $\kappa_m/\omega_m$   is on the order of $10^{-3}$\,\cite{Bourhill16,Wu21}.

If the estimated weak magnetic field $\vec B$ deviates from the bias field $\vec B_0$ along the $z$-axis, the Hamiltonian (\ref{H-RWA}) acquires an additional term $\!-\!(B_x\!-\!iB_y)\hat b/2\!-\!(B_x\!+\!iB_y)\hat b^\dagger/2$. 
In SM\,\cite{SM}, we show that these nonparallel components contribute to the displacement of the evolved Gaussian state and their impact on QFI is limited to the SNL. 
Significantly, the nonvanishing parallel component effectively encodes information into the covariance matrix, leading to QFI following HL.
Consequently, in our subsequent discussion, we  can safely disregard the nonparallel components\,\cite{SM}. 
Based on\,\cite{soykal:PRB2010}, crystalline anisotropy becomes notable in nanomagnets around 2nm radius. 
However, the YIG sphere considered here has a 360nm diameter\,\cite{Zhang14}. 
Thus, the effect of crystalline anisotropy can be disregarded.

The significance of dynamic entanglement in encoding information process becomes clearer through a contrasting example where the inputs are magnon and cavity squeezed coherent states with identical squeezing parameters. 
In this setup, no bipartite entanglement will be generated, indicating that information $B$ cannot be efficiently encoded into the phase parameter and is solely in the displacement parameter. 
Based on the analysis below Eq.\,(\ref{ent}), the precision cannot surpass the SNL, even in the presence of squeezing.

Returning to the initial scenario, however, entanglement during the measurement process will introduce thermalization, see Eq.\,(\ref{ent}).
Consequently, the initial squeezing resource $r_0$ cannot be  fully transferred  to the cavity,  disrupting  the  HL precision.   
Figure\,\ref{fig:fig2} (e) shows  the entanglement evolution concerning the detuning parameter $\omega_c-\omega_m$.
Vanishing entanglement in the curve  B (blue) indicates the quantum state periodically returns to the initial state. 
A nontrivial situation occurs in the resonant region A where entanglement vanishes after encoding the information into the cavity's covariance matrix, ensuring the realization of HL precision.
Far away from region A,  increasing detuning leads to strong entanglement between the cavity and the magnon, thus failing to suppress SNL, see Fig.\,\ref{fig:fig2} (a,c).

\begin{figure}[t]
	\includegraphics[width=\linewidth]{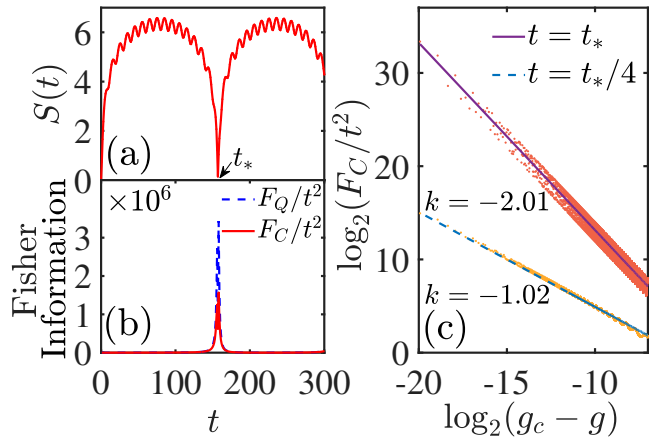}
	\caption{
Time evolution of (a) entanglement $S(t)$, (b) time-rescaled QFI $F_Q/t^2$, and CFI $F_C/t^2$ for critical dynamics.
(c) Scaling of CFI: $F_C\sim (g_c-g)^k$ with $k=-2.01$ for $t=t_*$ and $k=-1.02$ for $t=t_*/4$.
Lines in (c) are the results of fitting the theoretical data.
Other parameters are $\omega_c\!=\!\omega_m\!=\!2$, $g\!=\!0.9999g_c$, and  $t_*\!=\!\pi/\epsilon_-$.
}
	\label{fig:fig3}
\end{figure}

\emph{Critical dynamics.}--
In the strong coupling regime, Hamiltonian\,(\ref{H-1}) is directly diagonalizable as 
$\hat H\!=\!\epsilon_-\hat{c}^\dagger_1\hat{c}_1\!+\!\epsilon_+\hat{c}^\dagger_2\hat{c}_2$
using the Bogoliubov transformation\,\cite{Emary03}. 
The normal-to-superradiant phase transition occurs at the critical point $g_c\!=\!\sqrt{\omega_c\omega_m}/2$.
Near this point ($g\!\to\! g_c$), the excitation energy of the photon branch scales as $\epsilon_-\!\sim\!(g_c-g)^{1/2}$ and its derivative plays a crucial role in  achieving criticality-enhanced metrology.
Thus, in contrast to the weak coupling case, no specific resource state is required and we can choose the vacuum states as inputs.

Using Eqs.\,(\ref{d-CM-2-StandForm}, \ref{QFI-final}, \ref{CFI-final}), we derived the analytic expression of the covariance matrix $\gamma^c(t)$ in\,\cite{SM}.  
There is a  special time $t_*=n\pi/\epsilon_-$ $(n\in\mathbb Z_{>0})$ at which the covariance matrix $\gamma^c(t_*)$ is finite otherwise it diverges.
It then follows from Eq.\,(\ref{d-CM-2-StandForm}) that the divergence of $\gamma^c$ at $t\neq t_*$ implies the divergence of the number of thermalized particles $n_{\rm th}$, e.g.,
$n_{\rm th}(t_*/4)\!\sim\! (g_c-g)^{-1/2}$\,\cite{SM}.
Therefore, by the relation\,(\ref{ent}) we find that the magnon-photon entanglement $S(t)$ becomes large at $t\neq t_*$ but almost disappears suddenly at $t=t_*$, see Fig.\,\ref{fig:fig3}(a). 

Figure\,\ref{fig:fig3}(b) shows that the measurement precision has a maximum at the time $t_*$, meanwhile  the entanglement almost disappears at $t=t_*$. 
In this sense,  the vanishing entanglement in the measurement process enhances the measurement precision.
More rigorously, we  obtain critical scalings of the relevant parameters\,\cite{SM-2} and QFI:
\begin{flalign}\label{Critical-SC}
&F_Q(t_*)\sim  F_C(t_*)\sim (g_c-g)^{-2} t_*^2,\n 
&F_Q(t_*/4) \sim  F_C(t_*/4)\sim (g_c-g)^{-1} t_*^2,
\end{flalign}
which are further confirmed by numerical results shown in Fig.\,\ref{fig:fig3}(c).

It is worth noting that, compared with the $t_*$ case, the divergent squeezing $[\cosh(2r)\!\sim\!(g_c-g)^{-1/2}]$ in $t_*/4$ case causes no further enhancement in  measurement precision.
This is mainly because such a divergence in the covariance matrix also leads to a strong entanglement that  makes the cavity state  more insensitive to the estimated parameter, see the scalings of $r'$ and $\phi'$ in\,\cite{SM-2}.

\emph{Conclusion.}--
The usable   bipartite entanglement in accessible metrological schemes has been  established by linking Fisher information to entanglement. 
We have elaborated  on the significance of entanglement in the efficient encoding process  for high-precision quantum estimation, particularly starting from the initial  product state. 
While unveiling the adverse impact of entanglement on the  final measurement process. 
These findings enable us to design  quantum-enhanced estimations within an experimentally feasible cavity magnonics system. 
In particular, we have proposed an approach to realizing the precision of the HL  in the weak coupling regime.
Regarding strong coupling, we  have demonstrated that the criticality-enhanced metrology should  attribute  to the  criticality-induced parameter sensitivity rather than criticality-induced squeezing.
Our protocols show a potential  application of quantum metrology  through  current experimental cavity magnonic systems  \cite{Wolski20,Silaev23,Rao23,Golovchanskiy21-1,Kockum19,Boventer18,Golovchanskiy21,Kostylev16,You23}, enabling quantum-enhanced metrology  with or  without a particular squeezed initial state.
%

\begin{acknowledgments}
We thank J. Yang, X.-H. Wang and Y.-G. Su for valuable comments and suggestions.
This work was supported by the NSFC key grants No. 92365202 and  No. 12134015,  the NSFC grant No. 11874393 and No. 12121004.
HLS was supported by the European Commission through the H2020 QuantERA ERA-NET Cofund in Quantum Technologies project ``MENTA''.

The authors  QKW and HLS contributed equally to this work.  
\end{acknowledgments}

\clearpage\newpage
\setcounter{figure}{0}
\setcounter{table}{0}
\setcounter{equation}{0}
\def\thefigure{S\arabic{figure}}
\def\thetable{S\arabic{table}}
\def\theequation{S\arabic{equation}}
\setcounter{page}{1}
\pagestyle{plain}

\begin{widetext}

\section*{Supplemental Material for ``Quantum Metrology in Cavity Magnomechanics"}

\begin{center}
\noindent{Qing-Kun Wan, Hai-Long Shi, and Xi-Wen Guan }

\end{center}

\section{I. Hamiltonian of the Cavity magnomechanical system}
In pursuit of  self-consistency, we will first derive the Hamiltonian describing the cavity magnomechanical system by employing the standard Holstein-Primakoff transformation method \cite{Holstein40}, see also references \cite{Yuan22} and \cite{Yuan20}.
The Hamiltonian of the cavity magnomechanical system takes the form
$\hat H=\hat H_{\text{FM}}+\hat H_{\text{ph}}+\hat H_{\text{int}}$:
\begin{flalign}
&\hat H_{\text{FM}}=-J\sum_{\left\langle i,j\right\rangle }\vec {S_i}\cdot\vec S_j-\mu\sum_i(B_0\hat S_i^z+\vec B\cdot\vec S_i),\n
&\hat H_{\rm ph}=\frac{1}{2}\int(\epsilon_0\vec E_{\rm ph}^2+\mu_0\vec H_{\rm ph}^2)d\vec r,\quad
	\hat H_{\text{int}}=-\sum_i\vec S_i\cdot\vec H_{\rm ph},
\end{flalign}
where $\hat H_{\text{FM}}$, $\hat H_{\text{ph}}$, and $\hat H_{\text{int}}$ are the Hamiltonians for the ferromagnet material, the optical cavity, and their interaction, respectively.
Here, $J>0$ is the exchange constant;
$\left\langle i,j\right\rangle$ denotes the nearest neighbor spins;
$\mu$ is gyromagnetic ratio;
$B_0$ is a strong polarized magnetic field;
$\vec B=(B_x,B_y,B_z)$ is the estimated weak magnetic field;
$\vec E_p$ and $\vec H_p$ are electric field and magnetic field of the electromagnetic wave;
$\epsilon_0$ and $\mu_0$ are vacuum permittivity and susceptibility.

Using the Holstein-Primakoff transformation:
\begin{flalign}\label{H-P-SM}
	&\hat S_i^z=S-\hat b^\dagger_i\hat b_i,\quad
	\hat S_i^x=\sqrt{\frac{S}{2}}(\hat b_i+\hat b_i^\dagger),\quad
	\hat S_i^y=-i\sqrt{\frac{S}{2}}(\hat b_i-\hat b_i^\dagger),
\end{flalign}
and the Fourier transformation: 
\begin{flalign}\label{F-SM}
\hat b_i=1/\sqrt{N}\sum_k\exp(i\vec k\cdot\vec r_i)\hat b_k,
\end{flalign}
we can reformulate $\hat H_{\rm FM}$ in terms of the bosonic operators $\hat b$ and $\hat b^\dag$ as follows
\begin{flalign}
	&\hat H_{\text{FM}}=4JS\sum_k(1-\cos(kd))\hat b_k^\dagger\hat b_k
	+\mu(B_0+B_z)\hat b_k^\dagger\hat b_k
	-\mu\sqrt{\frac{S}{2}}(B_x-iB_y)\hat b_{k=0}
	-\mu\sqrt{\frac{S}{2}}(B_x+iB_y)\hat b^\dagger_{k=0},
\end{flalign}
where $d$ is lattice constant.
To quantize electromagnetic field, we introduce the vector potential operator:
\begin{flalign}\label{QED}
A(\vec r,t)=\sum_k\sqrt{\frac{\hbar}{2\omega_k\epsilon_0}}(\hat a_k\vec u_k(r)e^{-i\omega_kt}-\text{h.c.}).
\end{flalign}
Subsequently, utilizing the formulas 
$\vec E_{\rm ph}=-\frac{\partial}{\partial t}A(\vec r,t)$ and
$\vec H_{\rm ph}=\frac{1}{\mu_0}\nabla\times A$, we obtain
\begin{flalign}
	\hat H_{\text{ph}}=\sum_k\frac{\omega_k\hbar}{2}(\hat c_k^\dagger\hat c_k+\hat c_k\hat c_k^\dagger).
\end{flalign}
By substituting Eqs. (\ref{H-P-SM}), (\ref{F-SM}) and (\ref{QED}) into the Hamiltonian of the interaction,
we have
\begin{flalign}
	\hat H_{\text{int}}=g\sum_k(
	\hat b_k\hat c_k^\dagger
	+\hat b_k^\dagger\hat c_k
	+\hat b_k\hat c_{-k}
	+\hat b_k^\dagger\hat c_{-k}^\dagger),
\end{flalign}
where $g$ is the coupling constant. 
As the photon can only couple with the magnon around $k=0$, the total Hamiltonian simplifies to
\begin{flalign}\label{quantize}
	\hat H=
	\omega_c\hat c^\dagger\hat c
	+\omega_m\hat b^\dagger\hat b
	+g(\hat c+\hat c^\dagger)(\hat b+\hat b^\dagger)
	-\mu\frac{B_x-iB_y}{2}\hat b
	-\mu\frac{B_x+iB_y}{2}\hat b^\dagger,
\end{flalign}
where $\omega_m=\mu(B_0+B_z)$ and $B_x, B_y$ are the magnetic anisotropy.
We address the case of magnetic anisotropy only in Section III; otherwise, we assume  $B_x=B_y=0$.

\section{II. Fisher Information for general single-mode Gaussian States}
During the dynamical evolution, the cavity state $\hat\rho_c$ keeps in the Gaussian form and thus can be fully characterized by the displacement vector $\vec d^c=(\ex{\hat X_c},\ex{\hat P_c})$ and the covariance matrix 
\begin{eqnarray}\label{def-CM}
\gamma^c=\begin{pmatrix}
2\ex{\hat X_c^2}-2\ex{\hat X_c}^2&\ex{\{\hat X_c,\hat P_c\}}-2\ex{\hat X_c}\ex{\hat P_c}\\
\ex{\{\hat X_c,\hat P_c\}}-2\ex{\hat X_c}\ex{\hat P_c}& 2\ex{\hat P_c^2}-2\ex{\hat P_c}^2
\end{pmatrix},
\end{eqnarray}
where $\{\hat A,\hat B\}:=\hat A\hat B+\hat B\hat A$, $\ex{\hat A}:=\Tr(\hat\rho_c \hat A)$, $\hat X_c=(\hat c^\dag+\hat c)/\sqrt{2}$, and $\hat P_c=i(\hat c^\dag-\hat c)/\sqrt{2}$.
Furthermore,
arbitrary single-mode Gaussian state can be rewritten as a displaced squeezed thermal state \cite{Adam95}, i.e., 
\begin{eqnarray}\label{DST-state}
\hat \rho_c=\hat D(\alpha)\hat{S}(\zeta)\hat{\rho}_{\rm th}\hat{S}^\dagger(\zeta)\hat{D}^\dagger(\alpha),
\end{eqnarray}
where $\hat{D}(\alpha)=\exp(\alpha\hat c^\dagger-\bar{\alpha}\hat c)$ is the  displacement operator with $\alpha\in\mathbb C$, 
$\hat{S}(\zeta)=\exp(-\zeta\hat c^{\dagger 2}/2+\bar{\zeta}\hat c^2/2)$ is the squeezing operator with $\zeta=r\exp(i\phi)$ ($r\geq0$),
and $\hat{\rho}_{\rm th}=\sum_n P_n\ket{n}\bra{n}$ is a thermal state with $P_n=n_{\rm th}^n/(1+n_{\rm th})^{1+n}$ and $\ket{n}$ being the Fock states.
Substituting Eq. (\ref{DST-state}) into Eq. (\ref{def-CM}) we obtain
\begin{eqnarray}\label{gamma-standard state}
&&d_1^c=\sqrt{2}\text{Re}(\alpha),\quad d_2^c=\sqrt{2}\text{Im}(\alpha),\n 
&&\gamma^c_{11}=(2n_{\rm th}+1)[\cosh(2r)-\sinh(2r)\cos(\phi)],\n
&&\gamma^{c}_{22}=(2n_{\rm th}+1)[\cosh(2r)+\sinh(2r)\cos(\phi)],\n
&&\gamma^c_{12}=\gamma^c_{21}=-(2n_{\rm th}+1)\sinh(2r)\sin(\phi),
\end{eqnarray}
and then we can determine $(\alpha, r, \phi,n_{\rm th})$ in terms of the displacement vector $\vec d^c$ and the elements of the covariance matrix $\gamma^c$  as follows
\begin{eqnarray}\label{d-CM-2-StandForm}
&&\alpha=\frac{1}{\sqrt{2}}(d_1^c+id_2^c),\quad
n_{\rm th}=\frac{\sqrt{\det(\gamma^c)}-1}{2},\quad
r=\frac{1}{2}{\rm arcosh}\left(\frac{\Tr(\gamma^c)}{2\sqrt{\det(\gamma^c)}}\right)\geq 0,\ {\rm and}\ 
\tan\phi=\left(\frac{2\gamma_{12}^c}{\gamma_{11}^c-\gamma_{22}^c}\right).
\end{eqnarray}

Next, we will derive quantum Fisher information (QFI) and classical Fisher information (CFI) for single Gaussian states (\ref{DST-state}) and express them in terms of parameters $(\alpha, r, \phi,n_{\rm th})$.
The QFI is defined as
\begin{eqnarray}\label{QFI-1}
F_Q:=\Tr(\hat\rho_c\hat{L}^2),
\end{eqnarray}
where $\hat{L}$ is symmetric logarithmic derivative (SLD) operator.
By rewriting the cavity state as the standard form  (\ref{DST-state}) $\hat{\rho}_c=\sum_n P_n\ket{\Psi_n}\bra{\Psi_n}$ with $\ket{\Psi_n}=\hat{D}(\alpha)\hat{S}(\zeta)\ket{n}$ and $P_n=n_{\rm th}^n/(1+n_{\rm th})^{1+n}$,
the SLD operator is given by \cite{Paris09}
\begin{eqnarray}\label{SLD}
\hat{L}
&=&\sum_n\frac{\p_B P_n}{P_n}\ket{\Psi_n}\bra{\Psi_n}+2\sum_{n,m}\frac{P_n-P_m}{P_n+P_m}\ex{\Psi_m|\p_B\Psi_n}\ket{\Psi_m}\bra{\Psi_n},
\end{eqnarray}
where the parameter $B$ is to be estimated and $\partial_B:=\partial/\partial_B$.
By using the following formula \cite{Hall15}:
\begin{eqnarray}
\p_B\left(e^{\hat{O}}\right)&=&\int_0^1e^{s\hat{O}}\left(\p_B\hat{O}\right)e^{(1-s)\hat{O}}  ds ,
\end{eqnarray}
we obtain
\begin{eqnarray}\label{Useful-F-1}
&&\p_B\hat{D}(\alpha)=\left[\alpha'\hat c^\dagger-\bar{\alpha}'\hat c+\frac{1}{2}\alpha\bar{\alpha}'-\frac{1}{2}\bar{\alpha}\alpha'\right]\hat{D}(\alpha),\n
&&\p_B\hat{S}(\xi)=
\left[-\frac{\xi'}{2}\left(\left[\frac{\sinh(2r)}{4r}+\frac{1}{2}\right]\hat c^{\dagger2}+e^{-2i\phi}\left[\frac{\sinh(2r)}{4r}-\frac{1}{2}\right]\hat c^2+e^{-i\phi}\left[\frac{\cosh(2r)}{4r}-\frac{1}{4r}\right](\hat c^\dagger\hat c+\hat c\hat c^\dagger)\right)-h.c.\right]\hat{S}(\xi),\n 
\end{eqnarray}
where $'$ denotes $\partial_B$ and $\xi=r\exp(i\phi)$.
Using Eq. (\ref{Useful-F-1}) we obtain 
\begin{eqnarray}\label{E-1}
\ex{\Psi_m|\p_B\Psi_n}
&=&\ex{m|\hat S^\dag(\xi)\hat D^\dag(\alpha)([\p_B \hat D(\alpha)]\hat S(\xi)+\hat D(\alpha)[\p_B\hat S(\xi)])|n}\n 
&=&A_1\delta_{n,m-1}-\bar A_1\delta_{n,m+1}+A_2\delta_{n,m-2}-\bar A_2\delta_{n,m+2},
\end{eqnarray}
where 
\begin{eqnarray}\label{E-2}
A_1&=&\alpha'\cosh(r)+\bar\alpha'\exp(i\phi)\sinh(r),\quad A_2=-\frac{i\phi'}{4}\exp(i\phi)\sinh(2r)-\frac{r'}{2}\exp(i\phi).
\end{eqnarray}
Substituting Eqs. (\ref{E-1}), (\ref{E-2}), and $P_n=n_{\rm th}^n/(1+n_{\rm th})^{1+n}$ into Eq. (\ref{SLD}), we obtain the following expression of the SLD operator 
\begin{eqnarray}\label{SLD-2}
\hat{L}&=&\hat{D}(\alpha)\hat{S}(\xi)\left(\hat{L}_0+\hat{L}_1 +\hat{L}_2\right)\hat{S}^\dagger(\xi)\hat{D}^\dagger(\alpha),
\end{eqnarray}
where
\begin{eqnarray}
&&\hat{L}_0=\frac{\hat c^\dagger\hat c-n_{\rm th}}{n_{\rm th}(1+n_{\rm th})}n_{\rm th}',\n
&&\hat{L}_1=\frac{2}{1+2n_{\rm th}}\left([\alpha'\cosh(r)+\bar{\alpha}'e^{i\phi}\sinh(r)]\hat c^\dag+[\bar\alpha'\cosh(r)+\alpha'e^{-i\phi}\sinh(r)]\hat c\right),\n
&&\hat{L}_2=\frac{2(1+2n_{\rm th})}{1+2n_{\rm th}+2n_{\rm th}^2}
\left(\left[-\frac{i\phi'}{4}e^{i\phi}\sinh(2r)-\frac{r'}{2}e^{i\phi}\right]\hat c^{\dagger2}+\left[\frac{i\phi'}{4}e^{-i\phi}\sinh(2r)-\frac{r'}{2}e^{-i\phi}\right]\hat c^2\right).
\end{eqnarray}
Finally, by substituting Eq. (\ref{SLD-2}) into the definition of the QFI (\ref{QFI-1}) we obtain 
\begin{eqnarray}\label{QFI}
F_Q\label{QFI-final}
&=&\frac{n_{\rm th}'^2}{n_{\rm th}(1+n_{\rm th})}
+\frac{4}{2n_{\rm th}+1}\left[\alpha'\bar{\alpha}'\cosh(2r)+\text{Re}(\bar{\alpha}'^2e^{i\phi}) \sinh(2r)\right]\n
& &+\frac{(1+2n_{\rm th})^2}{2(1+2n_{\rm th}+2n^2_{\rm th})} \left[\sinh^2(2r)\phi'^2+4r'^2\right],
\end{eqnarray}
where the first term only depends on the thermalization parameter $n_{\rm th}$, the displacement parameter $\alpha$ only affects the second term, 
and the last term  is closely related with the squeezing parameter $\xi=re^{i\phi}$.

For the metrology scheme based on the Gaussian measurements, we  use CFI to evaluate its performance, which is defined as
\begin{eqnarray}\label{CFI}
F_C:=\int d^2\chi \frac{[\p_B p(\chi|B)]^2}{p(\chi|B)},
\end{eqnarray}
where $p(\chi|B)=\Tr[\hat\rho_c\hat \Pi(\chi)]$ is the probability of obtaining measurement outcome $\chi=(q,p)$ by performing the Gaussian measurements on  the cavity state $\hat \rho_c$.
A general Gaussian measurement is given by
\begin{eqnarray}
\hat{\Pi}(\chi)=\frac{1}{2\pi}\hat{D}(\chi)\hat \Pi_0\hat{D}^\dagger(\chi),
\end{eqnarray}
where $\hat\Pi_0$ is a density matrix of a single-mode Gaussian state and $\hat D(\chi)=\exp[i(p\hat X_c-q\hat P_c)]$ is the displacement operator.
Without losing generality, here we set  $\vec d^c=0$. The  the CFI (\ref{CFI}) can be rewritten as \cite{Monras06}
\begin{eqnarray}\label{CFI-2}
F_C=\frac{1}{2}\Tr[M(M^{-1})'M(M^{-1})'],
\end{eqnarray}
where the prime denotes the derivative of $M^{-1}=(\gamma^c+\gamma^0)$ w.r.t the weak magnetic field $B$ and $\gamma^0$ denotes the covariance matrix of the Gaussian state $\hat\Pi_0$.

It follows from Eq. (\ref{gamma-standard state}) that the covariance matrix of the cavity state can be  rewritten as
\begin{eqnarray}
\gamma^c&=&(2n_{\rm th}+1)R^T(\phi)T(r) R (\phi),
\end{eqnarray}
where 
\begin{eqnarray}
R(\phi)=\begin{pmatrix}
\cos(\phi/2)&\sin(\phi/2)\\-\sin(\phi/2)&\cos(\phi/2)
\end{pmatrix},
\quad 
T(r)=\begin{pmatrix}
e^{-2r}&0\\0& e^{2r}
\end{pmatrix}.
\end{eqnarray}
Notice that different choices of $\gamma^0$ determine different Gaussian measurements.
We restrict $\hat \Pi_0$ to a squeezed vacuum  state and parameterize its covariant matrix $\gamma^0$ in terms of $\psi$ and $s$, i.e., $\gamma^0=R^T(\psi)T(s)R(\psi)$.
Substituting these conditions into Eq. (\ref{CFI-2}) we obtain
\begin{eqnarray}\label{CFI-3}
F_C
=\frac{1}{2}\text{Tr}\left(M(M^{-1})'M(M^{-1})'\right)
=\frac{1}{2}\text{Tr}\left(\Gamma^{-1}\Sigma\Gamma^{-1}\Sigma\right),
\end{eqnarray}
where 
\begin{eqnarray}
\Gamma&=&
R(\phi)(\gamma^0+\gamma^c)R^T(\phi)\n 
&=&
R^T(\psi-\phi)T(s)R(\psi-\phi)+(2n_{\rm th}+1)T(r)\n 
&=&
[\sinh(2s)\sin(\phi-\psi)]\sigma_x
-[(2n_{\rm th}+1)\sinh(2r)+\cos(\phi-\psi)\sinh(2s)]\sigma_z\n
& &+[(2n_{\rm th}+1)\cosh(2r)+\cosh(2s)]\mathds{1}_2,\n
\Sigma&=&
R(\phi)(\gamma^0+\gamma^c)'R^T(\phi)\n 
&=&
2 n_{\rm th}'T(r)-2(2n_{\rm th}+1)r'\sigma_zT(r)+(2n_{\rm th}+1)\frac{\phi'}{2}[R^T(\pi)T(r)+T(r)R(\pi)]\n
&=&-(2n_{th}+1)\phi'\sinh(2r)\sigma_x
-2[n'_{\rm th}\sinh(2r)+(2n_{\rm th}+1)r'\cosh(2r)]\sigma_z\n
& &+2[n'_{\rm th}\cosh(2r)+(2n_{\rm th}+1)r'\sinh(2r)]\mathds{1}_2.
\end{eqnarray}
Here, $\sigma_i$ $(i=x,y,z)$ are the standard Pauli matrices and $\mathds{1}_2$ is the 2$\times 2$ identity matrix.
Denoting 
$\Gamma=\Gamma_x\sigma_x+\Gamma_z\sigma_z+\Gamma_0\mathds{1}_2$, 
$\Sigma=\Sigma_x\sigma_x+\Sigma_z\sigma_z+\Sigma_0\mathds{1}_2$ and substituting them into Eq. (\ref{CFI-3}) we have 
\begin{eqnarray}\label{CFI-4}
F_C&=&\frac{1}{2}\text{Tr}\left(\Gamma^{-1}\Sigma\Gamma^{-1}\Sigma\right)\n 
&=&\frac{\left(\Gamma_x\Sigma_x+\Gamma_z\Sigma_z-\Gamma_0\Sigma_0\right)^2+\left(\Gamma_x\Sigma_0-\Gamma_0\Sigma_x\right)^2+\left(\Gamma_z\Sigma_0-\Gamma_0\Sigma_z\right)^2-\left(\Gamma_z\Sigma_x-\Gamma_x\Sigma_z\right)^2}{\left(\Gamma_x^2+\Gamma_z^2-\Gamma_0^2\right)^2}.
\end{eqnarray}
Choose $\psi=\phi$ and $s=r$ then the CFI (\ref{CFI-4}) reduces to 
\begin{eqnarray}\label{CFI-final}
F_C
&=&\frac{[n'_{\rm th}-(2n_{\rm th}+1)r']^2}{2(n_{\rm th}+1)^2}
+\frac{[n'_{\rm th}+(2n_{\rm th}+1)r']^2}{2(n_{\rm th}+1)^2}
+\frac{(2n_{\rm th}+1)^2[\phi'\sinh(2r)]^2}{4(n_{\rm th}+1)^2}.
\end{eqnarray}

\section{III. Quantum Dynamics under the Rotating-Wave Approximation}

\subsection{Parallel Magnetic Field: $B_x=B_y=0$}

The Hamiltonian with a rotating-wave approximation (RWA) considered in the main text is given by 
\begin{eqnarray}\label{H-RWA}
\hat H=\omega_c\hat{c}^\dagger\hat{c}+\omega_{m}\hat{b}^\dagger\hat{b}+g\left( \hat{c}^\dagger\hat{b}+\hat{c}\hat{b}^\dagger\right),
\end{eqnarray}
where subscript $c$ and $m$ denote the cavity mode and the magnon mode, respectively.
The estimated parameter $B$  only exists in the frequency  $\omega_m$ of the magnon mode.
The dissipative dynamics of the system is described by the following quantum Langevin equation
\begin{eqnarray}\label{Leq-RWA}
&&\p_t\hat{c}(t)=-i\omega_c \hat{c}(t)-ig\hat{b}(t)-\frac{\kappa_c}{2}\hat{c}(t)+\sqrt{\kappa_c}\hat{c}_{\rm in}(t)\n
&&\p_t\hat{b}(t)=-i\omega_{m} \hat{b}(t)-ig\hat{c}(t)-\frac{\kappa_m}{2}\hat{b}(t)+\sqrt{\kappa_m}\hat{b}_{\rm in}(t),
\end{eqnarray}
where $\kappa_c$ and $\kappa_m$ represent the damping rate of the cavity mode  and the magnon mode, respectively.
The input noises are described by the annihilation operators $\hat c_{\rm in}$ and $\hat b_{\rm in}$, which have the relations $[\hat{c}_{\rm in}(t_1),\hat{c}^\dagger_{\rm in}(t_2)]=\delta(t_1-t_2)$, $[\hat{b}_{\rm in}(t_1),\hat{b}^\dagger_{\rm in}(t_2)]=\delta(t_1-t_2)$,
$\langle\hat{c}^\dagger_{\rm in}(t_1)\hat{c}_{\rm in}(t_2)\rangle=n_{c}\delta(t_1-t_2)$, and $\langle\hat{b}^\dagger_{\rm in}(t_1)\hat{b}_{\rm in}(t_2)\rangle=n_{m}\delta(t_1-t_2)$.
The thermal noise determine the values of $n_c$ and $n_m$, and for the low-temperature case we may choose $n_b=n_c\sim0$. 

We rewrite Eq. (\ref{Leq-RWA}) as 
\begin{eqnarray}\label{Leq-RWA-2}
\p_t
\begin{bmatrix}
\hat{c}(t)\\ i\hat{b}(t)
\end{bmatrix}=A
\begin{bmatrix}
\hat{c}(t)\\ i\hat{b}(t)
\end{bmatrix}+\vec V(t),
\end{eqnarray}
where 
\begin{eqnarray}
A=\begin{bmatrix}-i\omega_c-\frac{\kappa_c}{2}&-g\\g&-i\omega_{m}-\frac{\kappa_m}{2}\end{bmatrix},\quad 
\vec V(t)=\begin{bmatrix}
\sqrt{\kappa_c}\hat{c}_{\rm in}(t)\\
i\sqrt{\kappa_m}\hat{b}_{\rm in}(t)
\end{bmatrix}.
\end{eqnarray}
The solution of Eq. (\ref{Leq-RWA-2}) is thus given by
\begin{eqnarray}\label{solution}
\begin{bmatrix}
		\hat{c}(t)\\i\hat{b}(t)
\end{bmatrix}=T(t)
\begin{bmatrix}
		\hat{c}(0)\\i\hat{b}(0)
\end{bmatrix}+\int_{0}^{t}dsT(s)\vec V(t-s),
\end{eqnarray}
where
\begin{eqnarray}\label{T}
T(t)&=&e^{At}\\
&=&
\frac{1}{ig\Delta}
\begin{bmatrix}
g\left(\lambda_++i\omega_c+\frac{\kappa_c}{2}\right)e^{\lambda_- t}
-g\left(\lambda_-+i\omega_c+\frac{\kappa_c}{2}\right)e^{\lambda_+ t}
&
g^2(e^{\lambda_- t}-e^{\lambda_+ t})
\\
\left(\lambda_++i\omega_c+\frac{\kappa_c}{2}\right)\left(\lambda_-+i\omega_c+\frac{\kappa_c}{2}\right)(e^{\lambda_+ t}-e^{\lambda_- t})
&
g\left(\lambda_++i\omega_c+\frac{\kappa_c}{2}\right)e^{\lambda_+ t}-g\left(\lambda_-+i\omega_c+\frac{\kappa_c}{2}\right)e^{\lambda_- t}
\end{bmatrix},\nonumber
\end{eqnarray}
and 
$\lambda_\pm=i(-\omega_c-\omega_{m}\pm\Delta)/2-(\kappa_c+\kappa_m)/4$ are the eigenvalues of matrix $A$ with $\Delta=\sqrt{4g^2+[\omega_c-\omega_{m}-i(\kappa_c-\kappa_m)/2]^2}$.

Substituting Eqs. (\ref{solution}) and (\ref{T}) into Eq. (\ref{def-CM}), we obtain the covariance matrix for the cavity as follows
\begin{eqnarray}\label{Cavity-State-RWA-1}
\gamma^c(t)=\eta(t)\gamma^c_{\rm in}(t)+[1-\eta(t)](2n_c+1)\mathds{1}_2,
\end{eqnarray}
where $\eta(t)=\exp(-\kappa_ct)$ and $\gamma_{\rm in}^c$ is the covariance matrix of the cavity state $\hat \rho_c(t)$ for $\kappa_c=\kappa_m=n_c=n_m=0$:
\begin{eqnarray}
[\gamma_{\rm in}^c(t)]_{11}
&=&
1+\left(e^{-2r_0}-1\right)\frac{4g^2\sin^2(\frac{\Delta}{2}t)}{\Delta^2}\sin^2\left(\frac{\omega_c+\omega_{m}}{2}t\right)
+\left(e^{2r_0}-1\right)\frac{4g^2\sin^2(\frac{\Delta}{2}t)}{\Delta^2}\cos^2\left(\frac{\omega_c+\omega_{m}}{2}t\right),\n
\ [\gamma_{\rm in}^c(t)]_{22}
&=&
1+\left(e^{2r_0}-1\right)\frac{4g^2\sin^2(\frac{\Delta}{2}t)}{\Delta^2}\sin^2\left(\frac{\omega_c+\omega_{m}}{2}t\right)
+\left(e^{-2r_0}-1\right)\frac{4g^2\sin^2(\frac{\Delta}{2}t)}{\Delta^2}\cos^2\left(\frac{\omega_c+\omega_{m}}{2}t\right),\n
\ [\gamma_{\rm in}(t)]_{12}&=&-\sinh(2r_0)\frac{4g^2\sin^2(\frac{\Delta}{2}t)}{\Delta^2}\sin\left[(\omega_c+\omega_{m})t\right].
\end{eqnarray}
Here we assume $\kappa_m=\kappa_c$ and $n_m=n_c$.
The initial state is chosen as the magnon being a squeezed vacuum state and the cavity being the vacuum state, i.e., 
\begin{eqnarray}
\gamma(0)=\begin{bmatrix}
\gamma^c(0)&0\\
0&\gamma^m(0)
\end{bmatrix},
\quad
\gamma^c(0)=\begin{bmatrix}
1&0\\
0&1
\end{bmatrix},
\quad
\gamma^m(0)=\begin{bmatrix}
e^{-2r_0}&0\\
0&e^{2r_0}
\end{bmatrix}.
\end{eqnarray}
According to Eq. (\ref{d-CM-2-StandForm}), the parameters $(\alpha, r, \phi,n_{\rm th})$ of the state $\hat\rho_c(t)$  are given by 
\begin{eqnarray}
&&\alpha=0,\n 
&&\phi=\pi-(\omega_c+\omega_m)t,\n 
&& 
r=\frac{1}{2}\ln\left(\frac{[1-\eta(t)](1+2n_c)+\eta(t)(1+2n_{\rm in})e^{2r_{\rm in}}}{\sqrt{(\eta(t)(1+2n_{\rm in})+[1-\eta(t)](1+2n_{c}))^2+4\eta(t)[1-\eta(t)](1+2n_{\rm in})(1+2n_{c})\sinh^2(r_{\rm in})}}\right),\n
&&
n_{\rm th}=\frac{1}{2}\sqrt{(\eta(t)(1+2n_{\rm in})+[1-\eta(t)](1+2n_{c}))^2+4\eta(t)[1-\eta(t)](1+2n_{\rm in})(1+2n_{c})\sinh^2(r_{\rm in})}-\frac{1}{2},
\end{eqnarray}
where
\begin{eqnarray}
r_{\rm in}
&=&
\frac{1}{2}\ln\left(\frac{1-\xi(t)+\xi(t)e^{2r_0}}{\sqrt{1+4\xi(t)[1-\xi(t)]\sinh^2(r_0)}}\right),\n
n_{\rm in}
&=&
\frac{1}{2}\sqrt{1+4\xi(t)[1-\xi(t)]\sinh^2(r_0)}-\frac{1}{2},
\end{eqnarray}
and $\xi(t)=4g^2\sin^2(\Delta t/2)/\Delta^2$.

\subsection{Nonparallel Magnetic Field: $B_x\neq 0$ and $B_y\neq0$}
In this subsection, we examine the scenario where the estimated weak magnetic field $\vec B$ is misaligned with the bias field $\vec B_0$.
Under RWA, the Hamiltonian considered is  as follows
\begin{flalign}
	\hat H=\omega_c\hat{c}^\dagger\hat{c}+\omega_{m}\hat{b}^\dagger\hat{b}+g\left( \hat{c}^\dagger\hat{b}+\hat{c}\hat{b}^\dagger\right)
	-\mu\frac{B_x-iB_y}{2}\hat b-\mu\frac{B_x+iB_y}{2}\hat b^\dagger.
\end{flalign}
We observe that the presence of magnetic anisotropy $(B_x, B_y)$ introduces a new term $\vec V_m$ into the original dynamical equation (\ref{Leq-RWA-2}) :
\begin{eqnarray}
	\p_t
	\begin{bmatrix}
		\hat{c}(t)\\ i\hat{b}(t)
	\end{bmatrix}=A
	\begin{bmatrix}
		\hat{c}(t)\\ i\hat{b}(t)
	\end{bmatrix}+\vec V(t)+\vec V_m,
\end{eqnarray}
where 
$\vec V_m=\begin{bmatrix}0\\i\mu\frac{B_x+iB_y}{2}\end{bmatrix}$.
Consequently, the solution (\ref{solution}) acquires an additional term $\int_{0}^{t}dsT(s)\vec V_m$, given by
\begin{flalign}
	\int_{0}^{t}dsT(s)\vec V_m=
	\frac{\mu(B_x+iB_y)}{2\Delta}
	\begin{bmatrix}
		g[\frac{\exp(\lambda_-t)-1}{\lambda}_--\frac{\exp(\lambda_+t)-1}{\lambda_+}]\\
		(\lambda_++i\omega_c+\kappa_c/2)\frac{\exp(\lambda_+t)-1}{\lambda_+}
		-(\lambda_-+i\omega_c+\kappa_c/2)\frac{\exp(\lambda_-t)-1}{\lambda_-}
	\end{bmatrix}.
\end{flalign}
Fortunately, this term originating from the magnetic anisotropy does not change the covariance matrix $\gamma^c$ of the cavity, but introduces a nonzero displacement, given by
\begin{flalign}\label{alpha-magnetic anisotropy}
	\alpha=\frac{\mu g(B_x+iB_y)}{2\Delta}(\frac{\exp(\lambda_-t)-1}{\lambda_-}-\frac{\exp(\lambda_+t)-1}{\lambda_+}).
\end{flalign}

As discussed in the main text, the contribution of nonzero displacement to QFI is at most the SNL, i.e., $F_Q\sim N_{\alpha}^1$ where $N_{\alpha}$ represents the photon number generated by the displacement.
Furthermore, we also observe from Eq. (\ref{alpha-magnetic anisotropy}) that the displacement $\alpha$ is independent of the initial squeezing $r_0$, which provides the primary photon number $N_c(t_*)$.
Thus, as depicted in Fig. (\ref{fig:magneticanisotropy}), we notice that the QFI remains nearly constant in the case of a vertical magnetic field $B_z=0$. 
In contrast, if the parallel competent exists, i.e., $B_z\neq 0$, then the covariance matrix of the evolved cavity state remains unchanged, with the only modification being the replacement of $\omega_m=\mu(|\vec B|+B_0) $ with $\omega_m=\mu(B_z+B_0)$ in Eq. (\ref{Cavity-State-RWA-1}).
This modification introduces an overall factor in the QFI but does not impact its scaling behavior, even in the presence of noise, as illustrated by the blue solid line and the red dashed line in Fig. (\ref{fig:magneticanisotropy}).

\begin{figure}[h]
	\centering
	\includegraphics[width=0.5\linewidth]{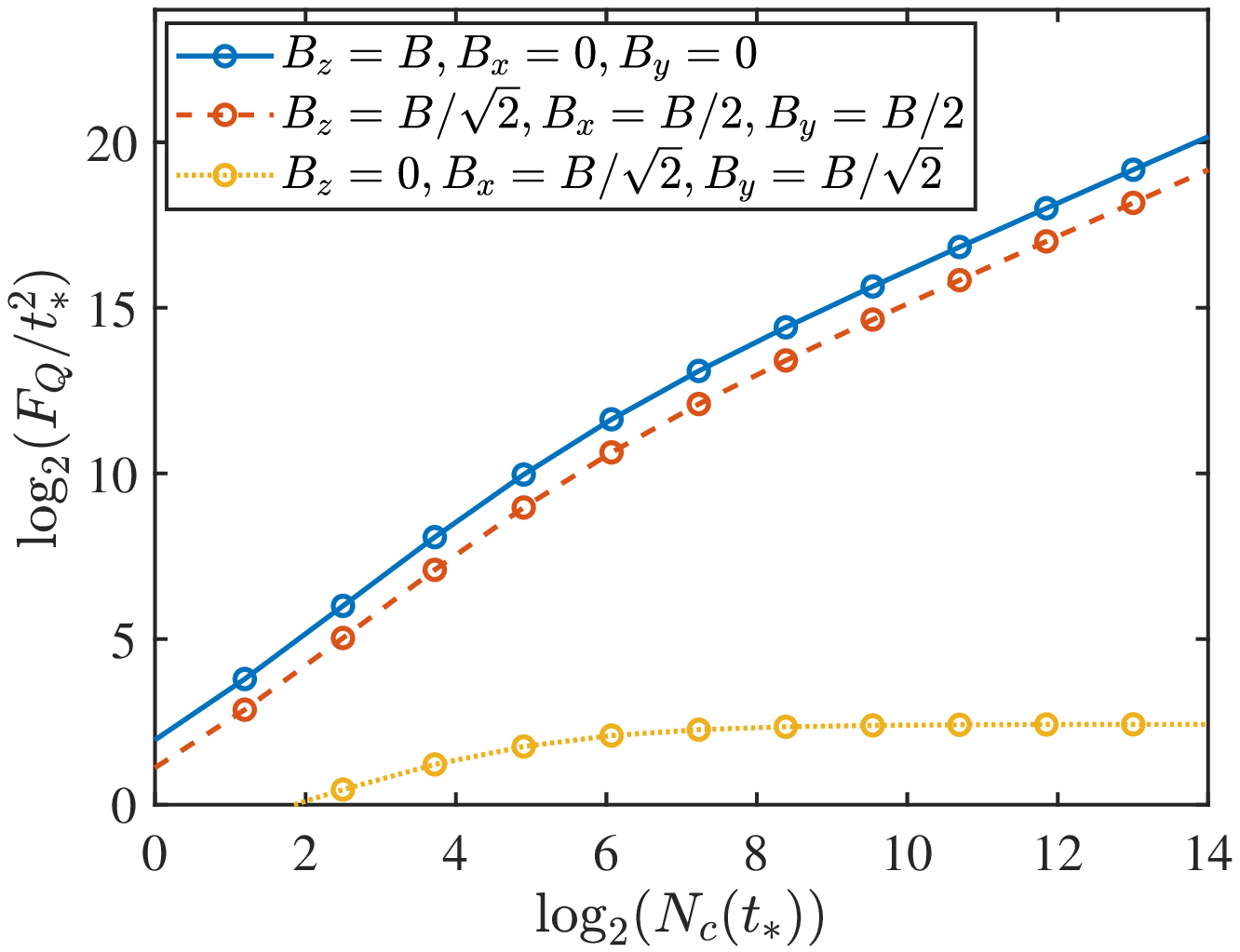}
	\caption{
	(Color online)
		The scaling relationships between the QFI and photon number $N_c$ are shown for different scenarios involving the estimated magnetic field $\vec B$: one where  $\vec B$ is completely polarized in the $z$-axis direction (blue solid-line), another where it is partially polarized in the $z$-axis direction (red dashed-line), and and a third where it is perpendicular to the $z$-axis direction (yellow dotted-line).
		In these simulations, we use parameters consistent with those employed in the experiment by \cite{Bourhill16}:
		$\omega_c=\omega_m=15.506\times2\pi$ GHZ,
		$g=7.11\times\pi$ GHZ,
		$\kappa_c=\kappa_m=1.029\times\pi$ MHZ.
		Additionally, the $t_*$ stands for the time corresponding to the point of maximum QFI and the thermal noise is given by $n_c=n_m=30$.
	}
	\label{fig:magneticanisotropy}
\end{figure}

\section{IV. QUantum Critical Dynamics (Beyond the RWA)}
For the study of quantum critical dynamics, we consider the following Hamiltonian including the  counter-rotating terms 
\begin{eqnarray}\label{H-beyondRWA-1}
\hat H=\omega_c\hat c^\dagger\hat c+\omega_m\hat{b}^\dagger\hat{b}+g(\hat c+\hat c^\dagger)(\hat{b}+\hat{b}^\dagger).
\end{eqnarray}
For $g<g_c=\sqrt{\omega_m\omega_c}/2$, the Hamiltonian (\ref{H-beyondRWA-1}) can be rewritten in terms of 
two uncoupled bosonic modes \cite{Emary03}:
\begin{eqnarray}
\hat H=\epsilon_-\hat{c}^\dagger_1\hat{c}_1+\epsilon_+\hat{c}^\dagger_2\hat{c}_2,
\end{eqnarray}
where $\epsilon_\pm^2=[\omega_c^2+\omega_m^2\pm\sqrt{(\omega_c^2-\omega_m^2)^2+16g^2\omega_c\omega_m}]/2$ and the relations between bosons $\{\hat c_1, \hat c_2\}$ and $\{\hat c, \hat b\}$ are
\begin{eqnarray}
	\begin{bmatrix}
		\hat c\\ \hat c^\dagger\\ \hat b\\ \hat b^\dagger
	\end{bmatrix}&=&\frac{1}{2}
	\begin{bmatrix}
		\frac{\cos(\delta)}{\sqrt{\omega_c\epsilon_-}}(\omega_c+\epsilon_-)
		&\frac{\cos(\delta)}{\sqrt{\omega_c\epsilon_-}}(\omega_c-\epsilon_-)
		&\frac{\sin(\delta)}{\sqrt{\omega_c\epsilon_+}}(\omega_c+\epsilon_+)
		&\frac{\sin(\delta)}{\sqrt{\omega_c\epsilon_+}}(\omega_c-\epsilon_+)\\
		\frac{\cos(\delta)}{\sqrt{\omega_c\epsilon_-}}(\omega_c-\epsilon_-)
		&\frac{\cos(\delta)}{\sqrt{\omega_c\epsilon_-}}(\omega_c+\epsilon_-)
		&\frac{\sin(\delta)}{\sqrt{\omega_c\epsilon_+}}(\omega_c-\epsilon_+)
		&\frac{\sin(\delta)}{\sqrt{\omega_c\epsilon_+}}(\omega_c+\epsilon_+)\\
		\frac{-\sin(\delta)}{\sqrt{\omega_0\epsilon_-}}(\omega_0+\epsilon_-)
		&\frac{-\sin(\delta)}{\sqrt{\omega_0\epsilon_-}}(\omega_0-\epsilon_-)
		&\frac{\cos(\delta)}{\sqrt{\omega_0\epsilon_+}}(\omega_0+\epsilon_+)
		&\frac{\cos(\delta)}{\sqrt{\omega_0\epsilon_+}}(\omega_0-\epsilon_+)\\
		\frac{-\sin(\delta)}{\sqrt{\omega_0\epsilon_-}}(\omega_0-\epsilon_-)
		&\frac{-\sin(\delta)}{\sqrt{\omega_0\epsilon_-}}(\omega_0+\epsilon_-)
		&\frac{\cos(\delta)}{\sqrt{\omega_0\epsilon_+}}(\omega_0-\epsilon_+)
		&\frac{\cos(\delta)}{\sqrt{\omega_0\epsilon_+}}(\omega_0+\epsilon_+)\\
	\end{bmatrix}
	\begin{bmatrix}
		\hat{c}_1\\\hat{c}_1^\dagger\\\hat{c}_2\\\hat{c}_2^\dagger
	\end{bmatrix}\equiv\hat{T}_2
	\begin{bmatrix}
		\hat{c}_1\\\hat{c}_1^\dagger\\\hat{c}_2\\\hat{c}_2^\dagger
	\end{bmatrix}.
\end{eqnarray}
Moreover, we define the operators $\hat T_1$, $\hat T_3(t)$ by
\begin{eqnarray}
	\begin{bmatrix}
		\hat{X}_c\\\hat{P}_c\\\hat{X}_m\\\hat{P}_m
	\end{bmatrix}&=&\frac{1}{\sqrt{2}}
	\begin{bmatrix}
		1&1&0&0\\
		-i&i&0&0\\
		0&0&1&1\\
		0&0&-i&i\\
	\end{bmatrix}
	\begin{bmatrix}
		\hat c\\ \hat c^\dagger\\ \hat b\\ \hat b^\dagger
	\end{bmatrix}\equiv \hat{T}_1
	\begin{bmatrix}
		\hat c\\ \hat c^\dagger\\ \hat b\\ \hat b^\dagger
	\end{bmatrix},\n
	\begin{bmatrix}
		\hat{c}_1(t)\\\hat{c}_1^\dagger(t)\\\hat{c}_2(t)\\\hat{c}_2^\dagger(t)
	\end{bmatrix}&=&
	\begin{bmatrix}
		\exp(-i\epsilon_- t)&0&0&0\\
		0&\exp(i\epsilon_- t)&0&0\\
		0&0&\exp(-i\epsilon_+ t)&0\\
		0&0&0&\exp(i\epsilon_+ t)\\
	\end{bmatrix}\equiv\hat{T}_3(t)
	\begin{bmatrix}
		\hat{c}_1(0)\\\hat{c}_1^\dagger(0)\\\hat{c}_2(0)\\\hat{c}_2^\dagger(0)
	\end{bmatrix},
\end{eqnarray}
where $\delta=\text{arctan}[4g\sqrt{\omega_c\omega_m}/(\omega_m^2-\omega_c^2)]/2$.
We consider the vacuum state as the initial state, i.e., $\gamma(0)=\mathds{1}_{4}$.
By the formular $\gamma(t)=\hat{M}(t)\gamma(0)\hat{M}^\dagger(t)$ with $\hat M(t)=\hat{T}_1\hat{T}_2\hat{T}_3(t)\hat{T}_2^{-1}\hat{T}_1^{-1}$, we obtain the covariance matrix of the cavity state as follows 
\begin{eqnarray}\label{rb}
[\gamma^c(t)]_{11}
&=&
1
+\cos^2\delta (\cos^2\delta-1)(\omega_c-\omega_m)
\left(\frac{2}{\omega_m} [\cos(\epsilon_+t)\cos(\epsilon_-t)-1]
-\frac{2\omega_c}{\epsilon_+\epsilon_-} \sin(\epsilon_+t)\sin(\epsilon_-t) \right)
\n 
& &-\frac{1}{2\epsilon_+^2\omega_m}(\cos^2\delta-1)\left[(\omega_c\omega_m+\epsilon_+^2)(\omega_c-\omega_m)\cos^2\delta+(\epsilon_+^2-\omega_c^2)\omega_m\right]\left[\cos(2\epsilon_+ t)-1\right]\n
& &-\frac{1}{2\epsilon_-^2\omega_m}\cos^2\delta\left[(\omega_c\omega_m+\epsilon_-^2)(\omega_c-\omega_m)\cos^2\delta+(\omega_m^2-\epsilon_-^2)\omega_c\right]\left[\cos(2\epsilon_- t)-1\right],
\n
\ [\gamma^c(t)]_{22}
&=&
1
-\cos^2\delta(\cos^2\delta-1)(\omega_c-\omega_m)
\left(\frac{2}{\omega_c}[\cos(\epsilon_+t)\cos(\epsilon_-t)-1]
-\frac{2\epsilon_+\epsilon_-}{\omega_c^2\omega_m}\sin(\epsilon_+t) \sin(\epsilon_-t)\right)
\n
& &+\frac{1}{2\omega_c^2\omega_m}(\cos^2\delta-1)\left[(\omega_c-\omega_m)(\omega_c\omega_m+\epsilon_+^2)\cos^2\delta+(\epsilon_+^2-\omega_c^2)\omega_m\right]\left[\cos(2\epsilon_+ t)-1\right]
\n
& &+\frac{1}{2\omega_c^2\omega_m}\cos^2\delta\left[(\omega_c-\omega_m)(\omega_c\omega_m+\epsilon_-^2)\cos^2\delta-(\epsilon_-^2-\omega_m^2)\omega_c\right]\left[\cos(2\epsilon_- t)-1\right],
\n
\ [\gamma^c(t)]_{12}
&=&
-\cos^2\delta(\cos^2\delta-1)(\omega_c-\omega_m)
\left(\frac{\epsilon_+^2+\omega_c\omega_m}{\epsilon_+\omega_c\omega_m}\sin(\epsilon_+t)\cos(\epsilon_-t)
+\frac{\epsilon_-^2+\omega_c\omega_m}{\epsilon_-\omega_c\omega_m}\cos(\epsilon_+t)\sin(\epsilon_-t) 
\right)
\n
& &+\frac{1}{2\epsilon_+\omega_c\omega_m}(\cos^2\delta-1)\left[\cos^2\delta(\omega_c-\omega_m)(\epsilon_+^2+\omega_c\omega_m)+(\epsilon_+^2-\omega_c^2)\omega_m\right]\sin(2\epsilon_+ t)
\n
& &+\frac{1}{2\epsilon_-\omega_c\omega_m}\cos^2\delta\left[\cos^2\delta(\omega_c-\omega_m)(\epsilon_-^2+\omega_c\omega_m)-(\epsilon_-^2-\omega_m^2)\omega_c\right]\sin(2\epsilon_- t).
\end{eqnarray}
Notice that $\cos^2\delta\neq 0,1$, $\epsilon_-=\sqrt{\frac{\omega_c^2+\omega_m^2}{2}\left(1-\sqrt{1+\frac{16\omega_c\omega_m}{(\omega_c^2+\omega_m^2)^2}(g^2-g_c^2)}\right)}\sim (g_c-g)^{1/2}\to0$ as $g\to g_c$ but $\epsilon_+$ is finite.
Thus from Eq. (\ref{rb}) we see that the finiteness of the covariance matrix $\gamma^c(t)$ requires $\sin(\epsilon_-t)=0$.
We denote such a time $t$ as $t_*:=n\pi/\epsilon_-$, $n\in\mathbb Z_{>0}$.
For time $t_*$ we deduce from Eq. (\ref{rb}) that the divergence of $[\gamma^c(t_*)]'$ behaves as follows
\begin{eqnarray}\label{t-star}
&&[\gamma^c(t_*)]_{11}'
\sim 
\left.\left[\frac{\sin(\epsilon_-t)}{\epsilon_-}\right]'\right|_{t=t_*}
\sim \frac{\epsilon_-'}{\epsilon}t_*\sim (g_c-g)^{-1}t_*, \n 
&&[\gamma^c(t_*)]'_{22}
\sim (g_c-g)^0t_*,\n 
&&[\gamma^c(t_*)]'_{12}
\sim \left.\left[\frac{\sin(\epsilon_-t)}{\epsilon_-}\right]'\right|_{t=t_*}
\sim \frac{\epsilon_-'}{\epsilon}t_*\sim (g_c-g)^{-1}t_*,
\end{eqnarray}
when $g\to g_c$.
Here we have used the result that $\epsilon'_-\sim(g_c-g)^{-1/2}$ and $\epsilon_-\sim (g_c-g)^{1/2}$ as $g\to g_c$.
Substituting Eq. (\ref{t-star}) into Eq. (\ref{d-CM-2-StandForm}) we obtain 
\begin{eqnarray}\label{n-r-phi-beyond-RWA}
&&n'_{\rm th}
=\frac{\det'(\gamma^c)}{4\sqrt{\det(\gamma^c)}}\sim(g_c-g)^{-1}t_*,\n 
&&\phi'=\frac{1}{\left(\frac{2\gamma^c_{12}}{\gamma^c_{11}-\gamma^c_{22}}\right)^2+1}\frac{2(\gamma^{c}_{12})'(\gamma^c_{11}-\gamma^c_{22})-2\gamma^c_{12}[\gamma^c_{11}-\gamma^c_{22}]'}{(\gamma^c_{11}-\gamma^c_{22})^2}\sim(g_c-g)^{-1}t_*,\n 
&&r'=\frac{1}{2\sqrt{\left(\frac{\Tr(\gamma^c)}{2\sqrt{\det(\gamma^c)}}\right)^2-1}}\frac{2\sqrt{\det(\gamma^c)}\Tr'(\gamma^c)-\det'(\gamma^c)\Tr(\gamma^c)/\sqrt{\det(\gamma^c)}}{4\det(\gamma^c)}\sim(g_c-g)^{-1}t_*.
\end{eqnarray}
Since the covariance matrix $\gamma^c(t_*)$ is finite then the criticality contribution to QFI and CFI both comes from the divergence of $r'$, $\phi'$ and $n'_{\rm th}$. 
By substituting Eq. (\ref{n-r-phi-beyond-RWA}) into Eqs. (\ref{QFI-final}) and  (\ref{CFI-final}) we obtain the scalings of QFI and CFI at time $t_*\sim(g_c-g)^{-1/2}$ as follows
\begin{eqnarray}
F_Q(t_*)\sim (g_c-g)^{-2}t_*^2,\n
F_C(t_*)\sim(g_c-g)^{-2}t_*^2. 
\end{eqnarray}

For $t\neq t_*$, the covariance matrix $\gamma^c$ is no longer finite but divergent as $g\to g_c$.
According to Eq. (\ref{rb}) we obtain 
\begin{eqnarray}\label{not-t-star-1}
&&[\gamma^c(t)]_{11}\sim \frac{1}{\epsilon_-^2}\sim(g_c-g)^{-1}\n
&&[\gamma^c(t)]_{22}\sim (g_c-g)^0\n 
&&[\gamma^c(t)]_{12}\sim\frac{1}{\epsilon_-}\sim(g_c-g)^{-1/2},
\end{eqnarray}
where $t\neq t_*$.
We further require that the time $t$ scales as $(g_c-g)^{-1/2}$ and makes $[\gamma^c(t)]'$ achieving its maximum divergence.
That is to say,
\begin{eqnarray}\label{not-t-star-2}
&&[\gamma^c(t)]_{11}'\sim\left[\frac{\cos(2\epsilon_-t)}{\epsilon_-^2}\right]'\sim (g_c-g)^{-3/2}t,\n  
\n  
&&[\gamma^c(t)]_{22}'\sim (g_c-g)^0t\n 
&&[\gamma^c(t)]_{12}'\sim\left[\frac{\sin(\epsilon_-t)}{\epsilon_-}\right]'\sim (g_c-g)^{-1}t,
\end{eqnarray}
which can be realized by choosing $t=t_*/4=\pi/(4\epsilon_-)$.
Substituting Eqs. (\ref{not-t-star-1}) and (\ref{not-t-star-2})
into Eq. (\ref{d-CM-2-StandForm}) we obtain 
\begin{eqnarray}\label{n-r-phi-beyond-RWA-2}
&&
n_{\rm th}=\frac{\sqrt{\det(\gamma^c)}-1}{2}\sim(g_c-g)^{-1/2} ,\n
&&
\cosh(2r)=\frac{\Tr(\gamma^c)}{2\sqrt{\det(\gamma^c)}}\sim(g_c-g)^{-1/2},\n
&&n'_{\rm th}
=\frac{\det'(\gamma^c)}{4\sqrt{\det(\gamma^c)}}\sim(g_c-g)^{-1}t_*,\n 
&&\phi'=\frac{1}{\left(\frac{2\gamma^c_{12}}{\gamma^c_{11}-\gamma^c_{22}}\right)^2+1}\frac{2(\gamma^{c}_{12})'(\gamma^c_{11}-\gamma^c_{22})-2\gamma^c_{12}[\gamma^c_{11}-\gamma^c_{22}]'}{(\gamma^c_{11}-\gamma^c_{22})^2}\sim(g_c-g)^0t_*,\n 
&&r'=\frac{1}{2\sqrt{\left(\frac{\Tr(\gamma^c)}{2\sqrt{\det(\gamma^c)}}\right)^2-1}}\frac{2\sqrt{\det(\gamma^c)}\Tr'(\gamma^c)-\det'(\gamma^c)\Tr(\gamma^c)/\sqrt{\det(\gamma^c)}}{4\det(\gamma^c)}\sim(g_c-g)^{-1/2}t_*.
\end{eqnarray}
Thus, by Eqs. (\ref{QFI-final}) and  (\ref{CFI-final}), we obtain the maximum scaling of QFI and CFI
\begin{eqnarray}
F_Q(t_*/4)\sim (g_c-g)^{-1}t_*^2,\n
F_C(t_*/4)\sim(g_c-g)^{-1}t_*^2,
\end{eqnarray}
when $t=t_*/4=\pi/(4\epsilon_-)$.

\bibliographystyle{IEEEtran}
\bibliography{Refer}

\begin{thebibliography}{}
\bibitem{FN-Gaussian}
 Here, $\hat{D}(\alpha)=\exp(\alpha\hat c^\dagger-\bar{\alpha}\hat c)$ is the  displacement operator, 
%
$\hat{S}(\zeta)=\exp(-\zeta\hat c^{\dagger 2}/2+\bar{\zeta}\hat c^2/2)$ is the squeezing operator with $\zeta=r\exp(i\phi)$,
%
and $\hat{\rho}_{\rm th}=\sum_n P_n\ket{n}\bra{n}$ is a thermal state with $P_n=n_{\rm th}^n/(1+n_{\rm th})^{1+n}$ and $\ket{n}$ is  the Fock state.
\bibitem{Footnote}
Note that $\gamma^c_{\rm sq}$ is the covariance matrix of the squeezed state $\hat S(-r_0e^{-i(\omega_c+\omega_m)t})\ket{0}$ in Eq. (\ref{Pro}).
\end{thebibliography}
\end{widetext}

\end{document}